\documentclass[conference]{IEEEtran}
\usepackage{cite}
\IEEEoverridecommandlockouts
\usepackage{epsfig}
\usepackage{amsmath}
\usepackage{theorem}
\usepackage{amsmath}
\usepackage{color}
\usepackage{mathrsfs}
\usepackage{amsfonts}
\usepackage[algo2e]{algorithm2e}
\usepackage{graphicx}
\usepackage{romannum}
\usepackage{multirow}
\usepackage{cite}
\usepackage{amsmath,amssymb,amsfonts}
\usepackage{algorithm}
\usepackage[noend]{algpseudocode}
\usepackage{graphicx}
\usepackage{textcomp}
\usepackage{amssymb}
\usepackage{mathrsfs}
\usepackage{euscript}
\usepackage{graphicx}
\usepackage{multirow}
\usepackage{cite}
\usepackage{amsmath,amssymb,amsfonts}
\usepackage{graphicx}
\usepackage{textcomp}
\usepackage{comment}
\def\BibTeX{{\rm B\kern-.05em{\sc i\kern-.025em b}\kern-.08em
    T\kern-.1667em\lower.7ex\hbox{E}\kern-.125emX}}
\begin{document}

\title{Energy and Spectral Efficiency Tradeoff in OFDMA Networks via Antenna Selection Strategy \\
}
\author{\IEEEauthorblockN {Ata Khalili$^{\dag}$~and Derrick Wing Kwan Ng$^{\star}$}
$^{\dag}$Department of Electrical and Computer Engineering,
Tarbiat Modares University, Tehran, Iran\\
$^{\star}$School of Electrical Engineering and Telecommunications,~The University of New South Wales, Australia\\
Email:~ata.khalili@ieee.org,~w.k.ng@unsw.edu.au
\thanks{D. W. K. Ng is supported by funding from the UNSW Digital Grid Futures Institute, UNSW, Sydney, under a cross-disciplinary fund scheme and by the Australian Research Council's Discovery Project (DP190101363).}}
\maketitle
\begin{abstract}
In this paper, we investigate the joint resource allocation and antenna selection algorithm design for uplink orthogonal frequency division multiple access (OFDMA) communication system.~We propose a multi-objective optimization framework to strike a balance between spectral efficiency (SE) and energy efficiency (EE).~The resource allocation design is formulated as
a multi-objective optimization problem (MOOP), where the conflicting objective functions are linearly combined into a single objective function employing the weighted sum method.~In order to develop an efficient solution, the majorization minimization (MM) approach is proposed where a surrogate function serves as a lower bound of the objective function. Then an iterative suboptimal algorithm is proposed to maximize the approximate objective function.~Numerical results unveil an interesting tradeoff between the considered conflicting system design objectives and reveal the improved EE and SE facilitated
by the proposed transmit antenna selection in OFDMA systems.
\end{abstract}
\vspace{-6mm}
\section{Introduction}
\vspace{-2mm}
 The next generation of wireless networks, namely 5G, aim at providing high data rate and system capacity.~To enable sustainable 5G networks, new communication technologies have been proposed to ameliorate the system energy efficiency (EE). In particular, various 5G techniques have been proposed which aim to enhance the network throughput while consuming less energy without sacrificing the quality of service \cite{1,Key}.~In practice, EE
maximization problem is more conspicuous in the uplink of wireless networks where extending the battery life of mobile users are of great significance. Moreover, reducing the transmit power
of users may also alleviate the interference power arising from co-channel users.~On the other hand,~in cellular networks, spectral efficiency (SE) is also an important performance measure. Moreover, nowadays users are employing multiple-antenna handsets while performing antenna selection can provide a higher flexibility to the system operator to strike a balance between SE and EE.

Recently, the studies of EE has drawn much attention for wireless orthogonal  frequency  division  multiple  access (OFDMA) communication
networks, where the available bandwidth is effectively divided into some orthogonal sub-channels
and each sub-channel can be individually assigned with a flexible manner \cite{3,4,5,6}. For instance,~in \cite{3}, the joint antenna-subcarrier-power allocation to maximize the EE of downlink multi-user was
studied where the Dinkelbach algorithm was proposed. However, to reduce the computational complexity of the algorithm, an upper bound on the maximum allowed interference is introduced as a constant to simplify the problem.~The joint
subcarrier assignment and power allocation of downlink transmission in two-tier heterogeneous
cellular networks was investigated in \cite{4} to maximize the system EE, where a two-step iterative solution and multi-objective
optimization problem (MOOP) based on epsilon constraint was proposed.
In \cite{6}, the authors demonstrated energy-efficient power allocation scheme for downlink multi-user distributed antenna systems in a single cell system. However,~sub-channel assignment for improving the system performance, as well as antenna selection, were not considered.

SE and EE are the key performance metrics for wireless communication systems, especially for
green communication networks~\cite{1}. The optimization of both these metrics can enhance
the system performance. Hence, it is important to take into account the non-trivial tradeoff between two
objectives in system design \cite{7,8,9}. In \cite{7}, the EE maximization in downlink multiuser
distributed antenna system was formulated as a MOOP which was solved via the weighting sum
method. The authors in \cite{8} investigated the tradeoff between EE and SE for a downlink single-cell and multi-cell scenario where the stochastic geometry approach was proposed to facilitate the analysis. In \cite{9}, the
authors formulated the resource allocation design as a MOOP which maximizes both the EE
and SE,~simultaneously. In addition, it was shown in \cite{9} that MOOP of EE-SE is equivalent to
maximizing data rate and minimizing the total power consumption, simultaneously.

To date, various low-complexity antenna selection methods have been developed \cite{10,11,12} to offer high flexibility to the system operator for throughput enhancement. In particular, antenna selection has been proposed for the uplink of 4G LTE-Advanced \cite{10} due to its low implementation cost and reduced feedback overhead compared to other techniques such as beamforming or precoding techniques.~Hence, the idea of antenna selection becomes an interesting research problem in various networks when the computational complexity of users is of great concern.~Although the full activation of multi-antenna systems enables a highly desirable gain,~simple antenna selection can extract a comparable gain of the multi-antenna systems with a marginal decrease of the performance~\cite{14,15,TWC_Ata}.

~
Motivated by the aforementioned observations, we investigate the problem of joint resource allocation and antenna selection in two scenarios. In the first scenario, the conventional antenna selection (CAS) selects only one antenna and only one RF chain is employed whereas, in the
second scenario, the generalized antenna selection (GAS) selects a subset of the antennas for
signal transmission \cite{TWC_Ata}.~We focus on a multicell system~and formulate a MOOP to
strike a balance between the SE and EE of a multi-cell uplink network, by considering QoS
requirement and feasibility of the transmit power level. The two competing objective functions
of this MOOP is handled by the weighting coefficient method which linearly combines them into a
single objective function. The resultant mixed integer non-linear problem (MINLP) is addressed
by using an efficient suboptimal algorithm with low computational complexity.~To facilitate the solution design, a penalty function is employed to handle the binary variable
constraints.~The underlying problem is
solved suboptimally via a method based on majorization minimization (MM) approach by constructing a
sequence of surrogate functions to iteratively approximate the non-convex objective function \cite{13}.

\section{System Model}\label{sec:System Model}
The uplink of a multi-carrier cellular network with the set of $\mathcal{B}=\{1,2,..,B\}$ cells is considered where the set of available sub-channel in each cell is represented by $\mathcal{S}=\{1,2...,S\}$\footnote{All BSs are equipped with multiple antennas. However, antenna selection
at BSs is out of scope of this work and is assumed to be predefined at the BSs, i.e., each antenna is reserved for a subset of users.}. The total number of users in each cell is ${M}$ where the set of user which is associated to BS b is denoted by $\mathcal{M}_{b}$ and each user is equipped with $\mathcal{Q}=\{1,2,...,Q\}$ antennas.~We assume that the perfect channel state information (CSI) is available at the resource allocator to design resource
allocation \cite{WCL_Ata}.~The uplink channel gain from the $m^{th}$ user in the $b^{th}$ cell to its desired BS, i.e., the $b^{th}$ BS, in the $s^{th}$ sub-channel from antenna $q$ is represented by $h^{sq}_{m{b},b}$.~Furthermore,~$a^{sq}_{mb}$ and $x^{s}_{mb}$ are defined as binary indicators showing if the $m^{th}$ user in the $b^{th}$ cell selects the $s^{th}$ sub-channel and the $q^{th}$ antenna. Accordingly, $p^{sq}_{mb}$ is the corresponding transmitted power from the $q^{th}$ antenna in the $s^{th}$ sub-channel for this user.


According to the Shannon capacity formula, the achievable rate of the $m^{th}$ user in the $b^{th}$ cell for the $s^{th}$ sub-channel, when the $q^{th}$ antenna is selected can be written as
\begin{equation}\label{sysmod2}
 \small R^{sq}_{mb}\triangleq\log_2\Bigg(1+\frac{p^{sq}_{mb}|h^{sq}_{m{b},b}|^2}{\sigma^2+\underset{b'\neq b, b' \in \mathcal{B}}{\overset{}{\mathop \sum }}~\underset{k\in \mathcal{M}_{b'}}{\overset{}{\mathop \sum }} \underset{q'\in \mathcal{Q}}{\overset{}{\mathop \sum }} a^{sq'}_{kb'} x^s_{kb'} p^{s{q'}}_{kb'} |h^{s{q'}}_{kb',b}|^2}\Bigg),\quad
\end{equation}\label{sysmod2}where $\sigma^2$ represents the additive white Gaussian noise~(AWGN) power.~Let us define the vector of power allocation,~sub-channel assignment,~and antenna selection variables as $\textbf{p}\in \mathbb{R}^{MBSQ\times 1} $,~$\textbf{x}\in \mathbb{Z}^{MBS\times 1}$,~and $\textbf{a}\in \mathbb{Z}^{MBSQ \times 1} $,~respectively.~In what follows we present antenna selection schemes for two scenarios, namely generalized antenna selection and conventional transmit antenna selection.
\subsection{Generalized Antenna Selection (GAS)}
As shown in Fig.~1,~for GAS approach, an OFDM symbol is transmitted from any antenna that the subcarriers are assigned to it.~The GAS\footnote{Through the problem formulation it is assumed that each sub-channel for each user can be assigned to one of the antennas and the
number of available RF chains is more than
one \cite{TWC_Ata}.} scheme aims to mitigate harmful effects of deep fades experienced in wireless channel by employing an optimal linear combining rule to a subset of the strongest available diversity branches,~thereby alleviating the receiver complexity and cost~\cite{14,15}.~In this case, a separate RF chain is required for each antenna.~In other words, a subset of the antenna must be selected.
\begin{figure} \label{fig1}
\centering
  \includegraphics[width=8.500cm,height=4.2500cm]{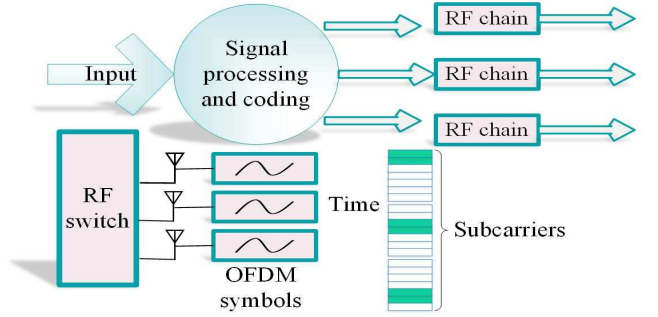}
  \caption{A block diagram of the generalized antenna selection scheme.}
\end{figure}
  ~The power consumption of the GAS is defined as $P^{c-GAS}_{m}=p^{\textrm{static}}_{m}+p^{\textrm{antenna}}_{m}\sum_{s}\sum_{q}a^{sq}_{mb}$ where $p^{\textrm {static}}_{m}$~is the static power required
 to support the basic circuit operations of the system at the transmitter and $p^{\textrm{antenna}}_{m}$ denotes the dissipated power per antenna \cite{Gold}.
\subsection{Conventional Antenna Selection (CAS)}
As shown in Fig.~2,~for CAS approach, a single transmit antenna is merely chosen and an OFDM symbol is transmitted via the selected antenna.~In fact, it is assumed that all sub-channels for each user can be assigned to one of the antennas and only one RF branch is available \cite{TWC_Ata}.~The model power consumption
of the CAS is presented as $P^{c-CAS}_{m}=p^{\textrm{static}}_{m}+p^{\textrm{antenna}}_{m}$.
In what follows, we study the resource allocation design for these two scenarios to strike a balance between EE and SE.
\begin{figure} \label{fig2}
\centering
  \includegraphics[width=8.2500cm,height=4.500cm]{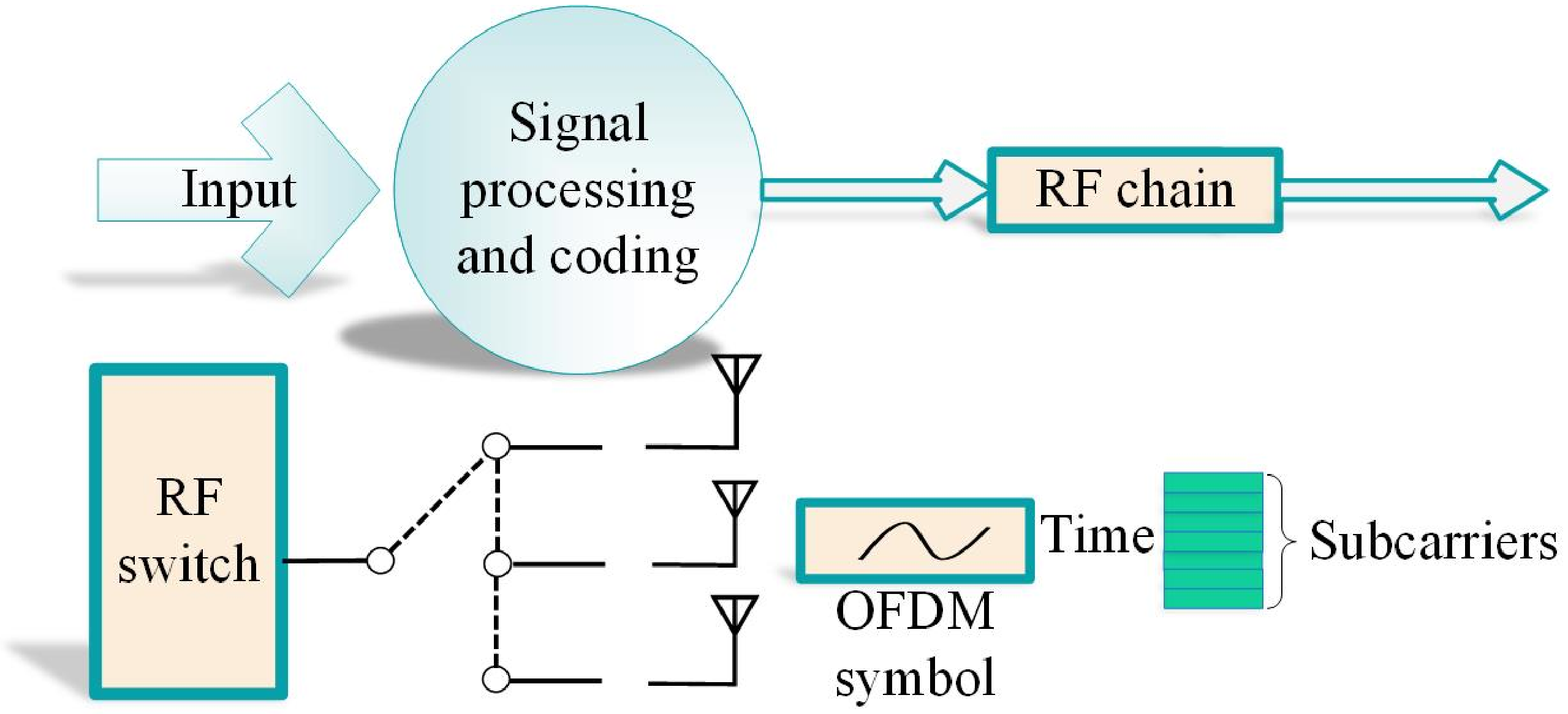}
  \caption{A block diagram of the conventional antenna selection scheme.}
 \end{figure}
\section{Problem Formulation }\label{sec:Problem Reformulation}
Energy efficiency (EE) of a wireless communication system is defined as a ratio between the total achievable rate and the total energy consumption given as follows:
\begin{align}\label{9}
\eta_{EE}=\frac{R^{\textrm{total}}}{P^{\textrm{total}}},
\end{align}
\vspace{-1mm}where $P^{\textrm{total}}=\frac{1}{\kappa} \underset{b\in\mathcal{B}}{\overset{}{\mathop \sum }}\underset{m\in \mathcal{M}_{b} }{\overset{}{\mathop \sum }}~\underset{s\in \mathcal{S}}{\overset{}{\mathop
\sum }}\underset{q\in\mathcal{Q}}{\overset{}{\mathop \sum }}\ \ p^{sq}_{mb}+\underset{b\in\mathcal{B}}{\overset{}{\mathop \sum }}\underset{m\in \mathcal{M}_{b} }{\overset{}{\mathop \sum }}P^{c}_{mb}$ and ${R}^{\textrm{total}}=\underset{b\in\mathcal{B}}{\overset{}{\mathop \sum }}\underset{m\in \mathcal{M}_{b} }{\overset{}{\mathop \sum }}~\underset{s\in \mathcal{S}}{\overset{}{\mathop
\sum }}\underset{q\in\mathcal{Q}}{\overset{}{\mathop \sum }} R^{sq}_{mb}$ are the total power consumption and data rate of the network,~respectively.~Note that  ${R}^{\textrm{total}}$,~is related to SE since we have ${R}^{\textrm{total}}=W\eta_{SE}$ (where $W$ represents the transmission bandwidth).~Moreover,~$P^{c}_{mb}$ denotes the consumed circuit power of the $m^{th}$ user device which depends on the CPU processing and includes the entire transmit RF chain which varies for the GAS and CAS scenarios.
Based on the configuration of the active links, the total circuit power consumption, $P^{c}_{mb}$, can be split into static and dynamic terms.~Also $0<\kappa<1$ denotes the efficiency of the power amplifier at each user.
As observed,~the power consumption of the two mentioned scenarios is different. For the GAS approach, all the assigned sub-channels can transmit a signal from different
antennas i.e., the GAS has more degrees of freedom to increase  the system throughput
as compared to the CAS scenario. However, in the GAS scenario the aggregate power
consumption increases with the number of activated RF chains. In
particular, the optimization of both scenarios can lead to a challenging tradeoff problem between
SE and EE.~In what follows, we aim at formulating two optimization problems for the two aforementioned scenarios to maximize both EE and SE through minimizing the inverse of EE and SE.
  \subsection{EE-SE for the GAS Scenario:}
The resource allocation and antenna selection for this scenario is formally stated as:
\begin{align}\label{10}
\mathcal{O}_{1}:~\min_{\textbf{a},\textbf{x},\textbf{p}}~&\eta_{EE}^{-1}(\textbf{a},\textbf{x},\textbf{p})\nonumber\\
\min_{\textbf{a},\textbf{x},\textbf{p}}~&\eta_{SE}^{-1}(\textbf{a},\textbf{x},\textbf{p})\nonumber\\
\text{s.t.}~&{\textrm C_1}:\underset{m\in \mathcal{M}_{b}
}{\overset{}{\mathop \sum }}x_{mb}^{s}=1,\nonumber\\
&{\textrm {C}_{2}}:\underset{q=1}{\overset{Q}{\mathop \sum
}}\,a_{mb}^{sq}=1,\nonumber\\
&{\textrm{C}_{3}}:p_{mb}^{sq}\ge 0,\\
&{\textrm{C}_{4}}:\underset{q\in{\mathcal{Q}}}{\overset{}
{\mathop \sum
}}\,\underset{s\in{\mathcal{S}}}{\overset{}{\mathop \sum
}}\,a_{mb}^{sq}x_{mb}^{s}p_{mb}^{sq}\le {{p}_{\max}},\nonumber\\
&{\textrm{C}_{5}}:~\underset{q\in{\mathcal{Q}}}{\overset{}
{\mathop \sum
}}\,\underset{s\in{\mathcal{S}}}{\overset{}{\mathop \sum
}}\,a_{mb}^{sq}x_{mb}^{s}R_{mb}^{sq}\ge {{R}_{\min}},\nonumber\\
&{\textrm{C}_{6}}:~x_{mb}^{s}\in \left\{ 0,1 \right\},~~\nonumber
\textrm C_7:a_{mb}^{sq}\in \left\{ 0,1 \right\}.\nonumber
\end{align}
In~$\mathcal{O}_{1}$,~constraints $\textrm C_1$ and $\textrm C_2$
 ensure each sub-channel in each cell can only be assigned to one user, and also, each user utilizes only one antenna in each sub-channel,~respectively.
~$\textrm C_3$ and $\textrm C_4$ indicate that the transmit power is non-negative and that the total transmit power of each user is limited to $p_{\max}$,~respectively.~$\textrm C_5$ guarantees
 a minimum data rate requirement for each user.~$\textrm C_6$ and
 $\textrm C_7$ indicate that the sub-channel and the antenna indicators take only binary values.\\
 \vspace{-4mm}
\subsection{EE-SE for the CAS Scenario:}
The resource allocation and antenna selection for this scenario can be written as:
\begin{align}\label{11}
\mathcal{O}_{2}:~\min_{\textbf{a},\textbf{x},\textbf{p}}~&\eta_{EE}^{-1}(\textbf{a},\textbf{x},\textbf{p})\nonumber\\
\min_{\textbf{a},\textbf{x},\textbf{p}}~&\eta_{SE}^{-1}(\textbf{a},\textbf{x},\textbf{p})\nonumber\\
\text{s.t.}~&{\textrm C_1}:\underset{m\in \mathcal{M}_{b}
}{\overset{}{\mathop \sum }}\underset{q\in \mathcal{Q}}{\overset{}{\mathop \sum }}x_{mb}^{sq}=1,\nonumber\\
&{\textrm {C}_{2}}:\underset{q=1}{\overset{Q}{\mathop \sum
}}\,a_{mb}^{q}=1,\nonumber\\
&{\textrm{C}_{3}}:p_{mb}^{sq}\ge 0,\\
&{\textrm{C}_{4}}:\underset{q\in{\mathcal{Q}}}{\overset{}
{\mathop \sum
}}\,\underset{s\in{\mathcal{S}}}{\overset{}{\mathop \sum
}}\,a_{mb}^{q}x_{mb}^{sq}p_{mb}^{sq}\le {{p}_{\max}},\nonumber\\
&{\textrm{C}_{5}}:~\underset{q\in{\mathcal{Q}}}{\overset{}
{\mathop \sum
}}\,\underset{s\in{\mathcal{S}}}{\overset{}{\mathop \sum
}}\,a_{mb}^{q}x_{mb}^{sq}R_{mb}^{sq}\ge {{R}_{\min}},\nonumber\\
&{\textrm{C}_{6}}:~x_{mb}^{sq}\in \left\{ 0,1 \right\},
\textrm C_7:a_{mb}^{q}\in \left\{ 0,1 \right\}\nonumber.
\end{align}
Constraints of $\mathcal{O}_{2}$ are similar to those of $\mathcal{O}_{1}$ except  $\textrm C_1$ and $\textrm C_2$.~Specifically,~$\textrm C_1$ indicates that each sub-channel inside each cell is assigned to at most one user and each user utilizes only one antenna for all its assigned sub-channels,~respectively.~Other constraints
 are similar to problem $\mathcal{O}_{1}$.~In the sequel,~we only focus on solving optimization $\mathcal{O}_{1}$ which is a general form of $\mathcal{O}_{2}$.~Furthermore, we compare the results obtained from GAS and CAS via numerical results and study the trade off between EE and SE for two aforementioned scenarios.~Now,~we propose a MOOP framework,~to strike a balance between EE and SE.~In order to handle the non-convex optimization problem $\mathcal{O}_{1}$,~we define a new MOOP that is a generalization of~$\mathcal{O}_{1}$.~It is proved that the optimization problem $\mathcal{O}_{1}$ can be equivalently written as the throughput is maximized while the aggregated power consumption is minimized simultaneously as follow~\cite{9}
 \begin{eqnarray}\label{12}
\begin{aligned}
  &\max_{\textbf{a},\textbf{x},\textbf{p}}\ R^{\textrm{total}}  \\
  &\min_{\textbf{a},\textbf{x},\textbf{p}}P^{\textrm{total}}~\text{s.t.}~{\textrm{C}_{1}}-{\textrm{C}_{7}}.
\end{aligned}
\end{eqnarray}
The weighting coefficients are employed  to convert the MOOP at hand into a SOOP that reflects the required preferences \cite{16}.~Therefore,~we rewrite the objective function of (\ref{12}) as:
\begin{equation}\label{13}
\begin{aligned}
~&\max_{\textbf{a},\textbf{x},\textbf{p}}~\textrm {EE-SE}=\frac{\nu}{w_{R}} {R}^{\textrm{total}}-\frac{(1-\nu)}{w_{P}}{P}^{\textrm{total}}\\
~&\text{s.t.}~{\textrm{C}_{1}}-{\textrm{C}_{7}},~~~
\end{aligned}
\end{equation}
where $w_{R}$,~and $w_{P}$~indicate the normalization factors.~Furthermore,~$0\leq\nu \leq 1$ denotes the weighting coefficient,~capturing the importance of different objectives (here EE and SE).
\vspace{-2mm}
 \section{Proposed solution}\label{sec:Problem Reformulation}
 \vspace{-2mm}
This section aims at providing a suboptimal solution for the problem  of  (\ref{13}). To this end, we first exploit the fact that $x^s_{mb}$ and $a^{sq}_{mb}$ take binary values at the optimal point.
Noting that the product term $a^{sq}_{mb} x^{s}_{mb}$ and the coupling with an affine variable $p^{sq}_{mb}$~in $\textrm{C}_{4}$ are the obstacles for the design of an efficient resource allocation algorithm.~By applying the big-M formulation,~we can rewrite $\textrm{C}_{4}$ into the following equivalent constraints:
  \begin{align}\label{sysmod4}
       \textrm{C}_{8}:~& p^{sq}_{mb}\leq x^{s}_{mb}~p_{\max},\nonumber\\
       \textrm{C}_{9}:~&p^{sq}_{mb} \leq a^{sq}_{mb}~p_{\max},\nonumber\\
       \textrm{C}_{10}:~& a^{sq}_{mb}\leq  x^{s}_{mb}.
  \end{align}
  In (7), it can be seen that $a^{sq}_{mb} x^{s}_{mb}$ is decoupled which simplifies the problem at hand.
Yet, even with equivalent constraint $\textrm{C}_{8}-\textrm{C}_{10}$ the problem (6) is still a MINLP which is complicated to solve. Thus, we take one step further to relax the binary constraints. The first step is to equivalently express the binary constraints $\textrm{C}_{6}$ and $\textrm{C}_{7}$  as the intersection of the following regions~\cite{17,18}
\begin{align}\label{sysmod6}
  &\mathcal{R}_1:0\leq a^{sq}_{mb}\leq 1, ~\,\mathcal{R}_2:0\leq x^s_{mb}\leq 1,\\
  &\mathcal{R}_3:\sum_{b}\sum_{m}\sum_{s}\sum_{q} \Big(a^{sq}_{mb}-(a^{sq}_{mb})^2+ x^s_{mb}- (x^s_{mb})^2\Big) \leq 0.\nonumber
\end{align}
For notational simplicity,~we define the feasible set spanned by all constraints and $R_{3}$~denoted by $\mathcal{D}$.~The optimization problem of (\ref{13}) can be written as follows:
\begin{eqnarray}\label{19}
\begin{aligned}
~&\max_{\textbf{a},\textbf{x},\textbf{p}}~\textrm {EE-SE}(\textbf{a},\textbf{x},\textbf{p})
~&\text{s.t.}~\textbf{a},\textbf{x},\textbf{p} \in \mathcal{D},~\mathcal{R}_1,~\mathcal{R}_2.
\end{aligned}
  \end{eqnarray}
   Then,~we propose a method to obtain an integer solution for $a^{sq}_{mb}$ and $x^{s}_{mb}$ as well as the power allocation policy.
To achieve this goal, we introduce a penalty term for the objective function in~(\ref{19}).~Thus, the problem is modified as:
\begin{eqnarray}\label{20}
  \max_{\textbf{a},\textbf{x},\textbf{p}}\ L(\textbf{a},\textbf{x},\textbf{p},\lambda) ~~~\text{s.t.}~~\textbf{a},\textbf{x},\textbf{p} \in \mathcal{D},\mathcal{R}_1,\mathcal{R}_2,
  \end{eqnarray}
where $L(\textbf{a},\textbf{x},\textbf{p},\lambda)$ is the \textit{abstract Lagrangian} of (\ref{19}), and is defined as
\begin{eqnarray}\label{21}
\begin{aligned}
 &L(\textbf{a},\textbf{x},\textbf{p},\lambda)\triangleq \textrm{EE-SE}(\textbf{a},\textbf{x},\textbf{p})\\-&\lambda\Bigg[ \sum_{b,m,s,q}\Big(a^{sq}_{mb} -(a^{sq}_{mb})^2\Big)+
 \sum_{b,m,s}\Big (x^{s}_{mb} -(x^{s}_{mb})^2\Big)\Bigg].
 \end{aligned}
\end{eqnarray}
Note that, $\lambda \gg 1$ acts as the penalty factor to penalize the objective function \cite{17,18,19}.~Furthermore,~we adopt a short hand notation $ \sum_{b,m,s,q}=\sum_{b\in \mathcal{B}}\sum_{m\in \mathcal{M}}\sum_{s\in \mathcal{S}}\sum_{q\in \mathcal{Q}}$
for the sake of presentation.

The optimization problem~(\ref{20})~is a continuous optimization problem with respect to all variables.~However,~due to the interference in the rate function,~the resulting optimization problem in~(\ref{20}) is still non convex.~To facilitate the solution design, we first rewrite the optimization problem
in the form of difference of concave (DC) function.~Mathematically speaking,~we have
\vspace{-2.5mm}
\begin{align}
&\frac{\nu}{w_{R}}\Big(F({\textbf{p}})-G({\textbf{p}})\Big)-\frac{(1-\nu)}{w_{P}}P^{\textrm{total}}\\ \nonumber&-\lambda\Big(D_{1}(\mathbf{a})-D_{2}(\mathbf{a})+E_{1}(\mathbf{x})-E_{2}(\mathbf{x})\Big),
\end{align}where
\vspace{-4.25mm}
 \begin{align}\label{sysmod24}
  &F({\textbf{p}})\!\! \triangleq\!\!\!\!\!\!\sum_{b,m,s,q}\!\! \Bigg(\log_2\Big({{p}^{sq}_{mb} |h^{sq}_{m{b},b}|^2}+\sigma^2 +\!\!\!\!\!\!\!\!\sum_{b'\neq b,k,q'}{p}^{sq'}_{kb'} |h^{sq'}_{k{b'},b}|^2\Big)\Bigg),\\
    &G({\textbf{p}})\triangleq \sum_{b,m,s,q} \Big(\log_2\big(\!\sigma^2 \!\!+\!\!\sum_{b'\neq b,k,q'}{p}^{sq'}_{kb'} |h^{sq'}_{k{b'},b}|^2)\Big),\\
 &D_{1}(\mathbf{a})\triangleq \sum_{b,m,s,q}\Big(a^{sq}_{mb}\Big),~D_{2}(\mathbf{a})\triangleq \sum_{b,m,s,q} \Big(a^{sq}_{mb}\Big)^{2},\\
&E_{1}(\mathbf{x})\triangleq \sum_{b,m,s}\Big(x^{s}_{mb}\Big), E_{2}(\mathbf{x})\triangleq \sum_{b,m,s}\Big(x^{s}_{mb}\Big)^{2}.
\end{align}
One can readily observe that the objective function can be written as the difference of two concave functions.
In a similar way, the left hand side of the constraint $\textrm{C}_{5}$ can be written as:
\begin{equation}\label{sysmod29}
  f_{mb}({\textbf{p}})-g_{mb}({\textbf{p}}),
  \end{equation}
  \vspace{-3mm}
where
\begin{align}\label{fsmal}
\small &f_{mb}({\textbf{p}})\! \triangleq \sum _{s,q}\log_2\bigg({{p}^{sq}_{mb}} |h^{sq}_{m{b},b}|^2+\sigma^2+\!\!\!\!\!\!\!\!\sum_{b'\neq b,k,q'}{p}^{sq'}_{kb'} |h^{sq'}_{k{b'},b}|^2\bigg),\\
 \small &g_{mb}({\textbf{p}})\triangleq \sum_{s,q} \log_2\bigg(\!\sigma^2 \!\!+\!\!\sum_{b'\neq b ,k,q'}{p}^{sq'}_{kb'} |h^{sq'}_{k{b'},b}|^2\bigg).\quad\quad
\end{align}
To solve the equivalent DC problem we apply the majorization minimization approach~\cite{13}~to obtain a locally optimal solution of~(12). To this end, the first order Taylor approximations for $G({\textbf{p}}),~D_{2}(\textbf{a}),~E_{2}(\textbf{x})$,~and $g({\textbf{p}})$ are being exploited to convert the DC problem into a convex problem which can be effectively solved iteratively.~Since $G({\textbf{p}}),~D_{2}(\textbf{a}),~E_{2}(\textbf{x})$,~and $g({\textbf{p}})$ are differentiable convex functions,~for any feasible point $\textbf{p}^{i}$,~$\textbf{a}^{i}$,~and $\textbf{x}^{i}$,~we have
\begin{align}
    &{G}({\textbf{p}})\leq G({\textbf{p}}^{i-1})
 +\nabla_{{\textbf{p}}}G^{T}({\textbf{p}}^{i-1})({\textbf{p}}-{\textbf{p}}^{i-1})\triangleq \tilde{G}({\textbf{p}}),\\
 &{D}_{2}(\textbf{a})\geq D_{2}(\textbf{a}^{i-1})
 +\nabla_{\textbf{a}}D_{2}^{T}(\textbf{a}^{i-1})(\textbf{a}-\textbf{a}^{i-1})\triangleq \tilde{D}_{2}(\textbf{a}),\\
  &{E}_{2}(\textbf{x})\geq E_{2}(\textbf{x}^{i-1})
 +\nabla_{\textbf{x}}E_{2}^{T}(\textbf{x}^{i-1})(\textbf{x}-\textbf{x}^{i-1})\triangleq \tilde{E}_{2}(\textbf{x}) ,\\
 &{g}_{mb}(\textbf{p})\leq g_{mb}({\textbf{p}}^{i-1})+\nabla_{{\textbf{p}}}g^T_{mb}({\textbf{p}}^{i-1})({\textbf{p}}- {\textbf{p}}^{i-1})\triangleq \tilde{g}_{mb}(\textbf{p}).
\end{align}\vspace{-1mm}Hence, for the $i^{th}$ iteration, noting $\tilde{G}({\textbf{p}})$,~$\tilde{E}_{2}(\textbf{a}),~D_{2}(\textbf{x})$, and $\tilde{g}_{mb}({\textbf{p}})$ are all affine functions.~Therefore,~a suboptimal solution of (12) can be obtained by solving the following convex optimization problem:
\begin{align}\label{sysmod37}
 \max_{\textbf{a},\textbf{x},{\textbf{p}}}
&\frac{\nu}{w_{R}}\Big(F({\textbf{p}})-\tilde{G}({\textbf{p}})\Big)-\frac{(1-\nu)}{w_{P}}P^{\textrm{total}}\nonumber\\ &-\lambda\Big(D_{1}(\mathbf{a})-\tilde{D}_{2}(\mathbf{a})+E_{1}(\mathbf{x})-\tilde{E}_{2}(\mathbf{x})\Big)\nonumber\\
   & \text{s.t.}\quad
    f_{mb}({\textbf{p}})-\tilde{g}_{mb}({\textbf{p}})\geq R_{\min},\nonumber\\
    &x^{s}_{mb},a^{sq}_{mb}\in [0,1],
     ~\textrm{C}_{1}-\textrm{C}_{4},~\textrm{C}_{8}-\textrm{C}_{10}.
     \end{align}
     The convex optimization problem of~(\ref{sysmod37}) is solved based on method as presented in \textbf{Algorithm 1}.\\
      \begin{algorithm}[t]\label{algorithm3}
 \caption{\small Proposed Method MOOP based on Successive Convex Approximation~(MOOP DC Programming)}
 \begin{algorithmic}[1]
  \State~Initialize $i=0$ and maximum number of iteration $I_{\max}$, penalty factor $\lambda\gg1$ , set appropriate weighting coefficient factor $(\nu)$ and feasible set vector $\textbf{p}^{0} $,~$\textbf{x}^{0}$, and $\textbf{a}^{0}$
\State {\textbf{~Repeat}}
\State~Update $\tilde{G}({\textbf{p}})$,~$\tilde{D}_{2}(\textbf{a})$,~$\tilde{E}_{2}(\textbf{x})$,~and $\tilde{g}({\textbf{p}})$ as presented in~({20}),~({21}),~({22}),~and~({23}),~respectively.
\State Solve optimization problem of ({24}) and store the intermediate resource allocation policy~$\textbf{a}^{i}$,~${\textbf{x}^{i}}$, and $\textbf{p}^{i}$
\State Set $i=i+1$ and $\textbf{p}^{i}=\textbf{p}$,~$\textbf{x}^{i}=\textbf{x}$,~and $\textbf{a}^{i}=\textbf{a}$
\State \textbf{~Until }convergence or $i=I_{\max}$
\State $\textbf{a}^{*}=\textbf{a}^{i}$,~$\textbf{x}^{*}=\textbf{x}^{i}$,~$\textbf{p}^{*}=\textbf{p}^{i}$
\end{algorithmic}
\end{algorithm}
\vspace{-4mm}
       \section{Computational Complexity Analysis }\label{complexity}
       \vspace{-2mm}
In this section, we analyze the computational complexity
of considered optimization problems in the
proposed algorithm in \textbf{Algorithm 1} via DC programming \cite{17,22}.
For the optimization problem there are totally $BMSQ$ decision variables and $SB+BMS+BM+BM+BMS+2MBSQ$ convex and affine constraints in the convex program (\ref{10}).~Therefore,~solving the optimization problem via the interior point method requires the order of  $\mathcal{O}(BMSQ)^{3}(SB+2BM+2BMS+2MBSQ)$ which is polynomial time complexity.
~An exhaustive search would require the examination
of all $(M)^{BSQ}B^{Q}B^{S}$ possible sub-channel assignment and
antenna selection choices. Note that exhaustively searching over
all possible choices of sub-channels/antennas is a time consuming
task even for moderate number of antennas/users and sub-channels.~Furthermore, when CVX is employed to solve the problems in (24),~it employs D.C. with the interior point method and the number of required iterations for this approach is $\frac{\log(c)/t^{0}\delta}{\log(\epsilon)}$,~where $c$ is the total number of constraints,~$t^{0}$ is the initial point,~$0<\delta\ll 1$~is the stopping criterion,~and $\epsilon$ is used for updating the accuracy of the method~\cite{AH,22,TVT_Ata}.
\vspace{-2mm}
\section{Numerical Results}\label{sec:sim}
\vspace{-2mm}
In this section, the performance of our proposed approach is investigated.~We adopt the simulation parameters given in Table I, unless
specified otherwise.~The wireless channel gains are Rayleigh fading set as $h^{n}_{mb}=\phi^{n}d^{-\alpha}_{mb}$, where $d_{mb}$ is the distance between users and the base stations. The carrier frequency and sub-channel bandwidth are 2 GHz and 180 kHz, respectively.~Monte Carlo
simulation is performed by generating random channel realizations and we compute the average system performance.
\begin{table}
  \centering
\caption{simulation parameters}
\label{Simulation Parameters}
\begin{tabular}{|c|c|}\hline
{\bf Parameter} & {\bf Value} \\ \hline \hline
{Cell radius} & $500$ m \\
The number of cell & $4$\\
The number of user in each cell & $4$\\
 $N$,~subcarrier bandwidth  & \{$8,16$\} and {$180$} kHz \\
Noise power ($N_0$) & {$-174$} dBm/Hz \\
{Path loss exponent ($\alpha$)} & {$3$} \\
 {$P_{m}^\textnormal{static}$} & {$10$ dBm} \\
 {$P_{m}^\textnormal{antenna}$} & {$7$ dBm} \\
$p_{\textrm{max}}$ & {$23$ dBm} \\
$R_{\textrm{min}}$ & $5$ bps/Hz \\
Number of channel realization & $100$\\
\hline
\end{tabular}
\end{table}
 \begin{figure}[t]
\centering
\includegraphics[width=10.00cm ,height=5.00cm]{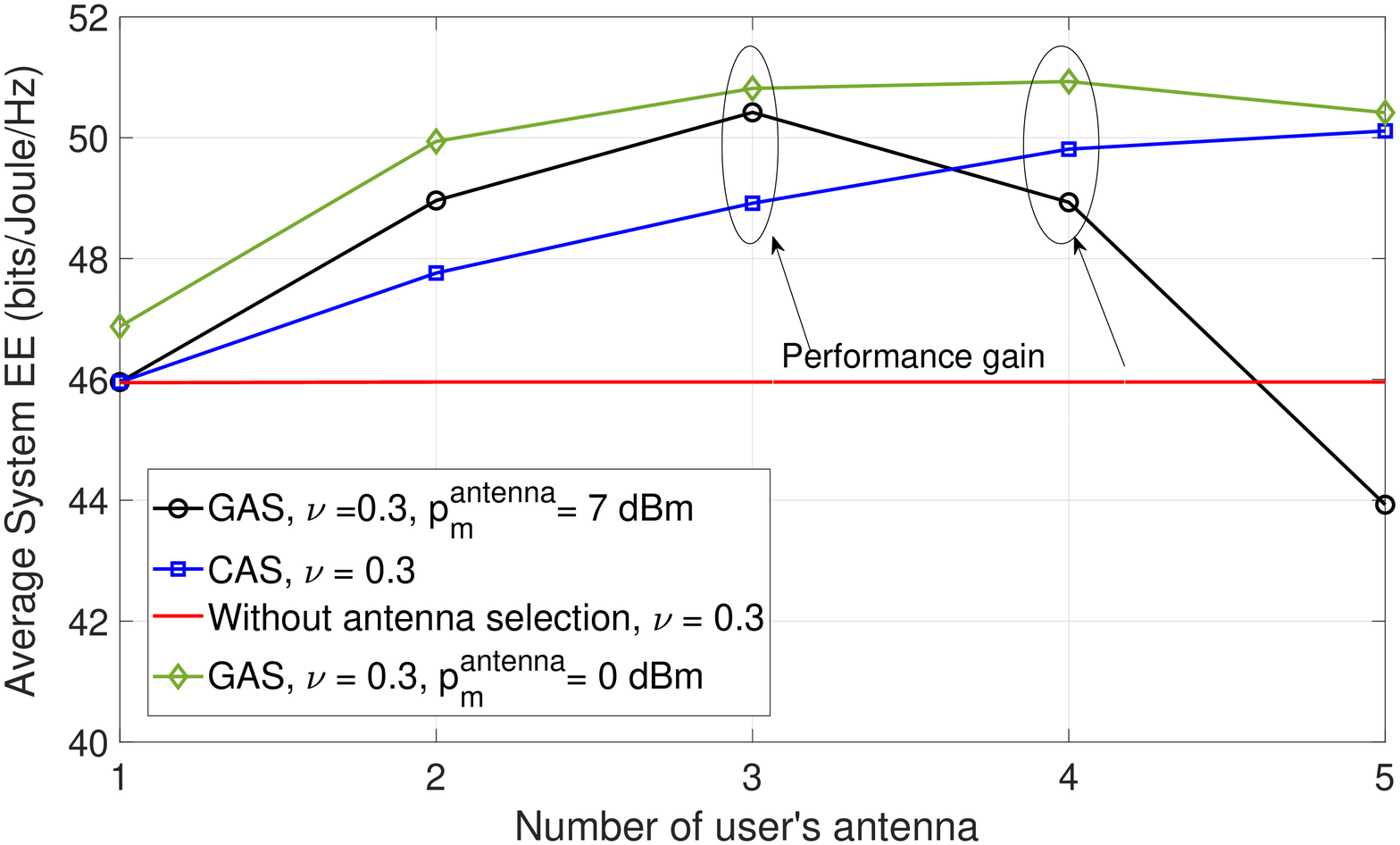}
\vspace{-6mm}
  \caption{ Average system EE versus different number of user's antenna for the three schemes.}
  \label{fig:ee_antenna}
\end{figure}
\begin{figure}[t]
\centering
\includegraphics[width=9.00cm ,height=5.00cm]{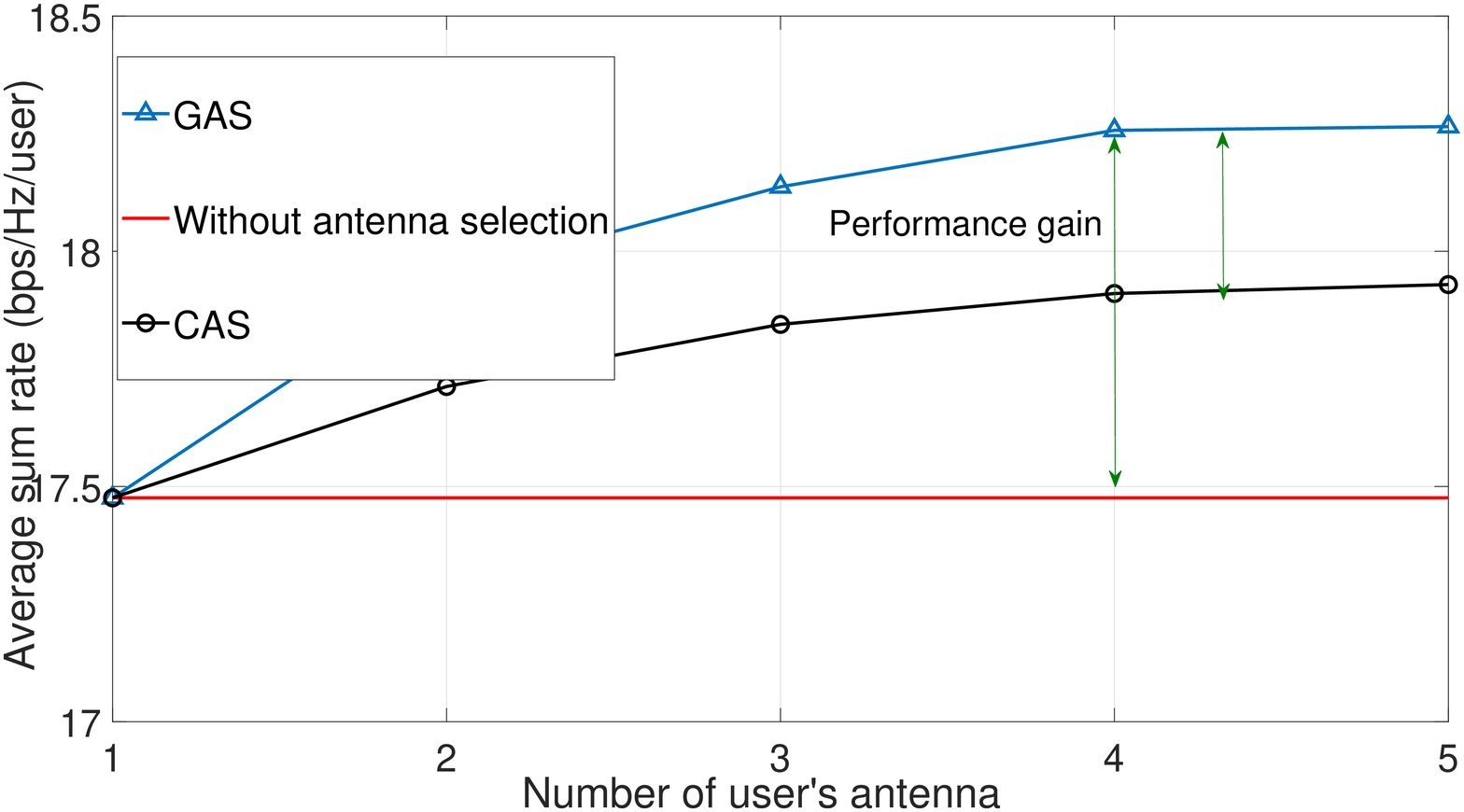}
\vspace{-6mm}
\caption{Average sum rate versus different number of user's antenna for the three schemes.}
  \label{fig:ee_antenna}
\end{figure}
 Fig.~3 and Fig.~4 illustrate the EE and SE versus number of user's antenna for different schemes namely,~GAS,~CAS,~and without antenna selection scenarios,~respectively.
 It can be concluded from Fig.~3 that in the GAS scenario for a large value of dynamic power, by increasing the number of user's antenna, the EE first increases and reaches to a maximum value and then decreases with the number of user's antenna.~This is because when the number of user's antenna increases, the dynamic term of the power consumption grows linearly which leads to a reduction of EE.~Moreover,~from this figure, we observe that the EE increases as the dynamic power decreases.~We also observe that the slope of the curve decreases with the number of user's antennas.~However, the slope of the curve  varies from one case to another.~For instance, for the GAS scenario with $P_{m}^\textnormal{antenna}=0$ dBm, the slope of the curve starts to decline for $Q>4$ while for GAS scenario with $P_{m}^\textnormal{antenna}=7$ dBm  starts to decrease when $Q>3$.~The reason for this trend is that~once the maximum EE of the system is achieved,~a further increase in the number of user's antenna would result in a degradation in EE due to increasing the dynamic term of power consumption.

 On the other hand,~as illustrated in Fig.~4 in the GAS scenario, the system sum rate improves by increasing the number of user's antenna.
  It should be noted that although GAS is an effective method to improve the systems throughput, it increases the power consumption of the network.
This is because a subset of the antenna,~must be selected which increases the power consumption of user per RF
chain.
Moreover as illustrated in Fig.~3, for the CAS scenario when the number of user's antenna increases,~the EE increases monotonically
and becomes saturated.~As shown in Fig.~4,~although CAS improves system throughput~(in
comparison to the scenario where no antenna selection scheme is employed), its data rate is still less than that of GAS. On the other hand, the aggregated power consumption of CAS is less than GAS. This is due to the fact in CAS, we have only one RF chain which reduces the dynamic term of the power consumption model,~leading to EE enhancement~(in comparison to the GAS scenario).
In fact, there is a non-trivial trade off between the data rate of the system and its total power consumption.
\vspace{-3mm}
\section{Conclusion}\label{conclusion}
\vspace{-1.5mm}
This paper investigated the tradeoff between EE and SE of uplink multi-cell networks via joint
sub-channel assignment, power control, and antenna selection for two scenarios. In the first scenario, known as CAS, there is only one RF chain available which all sub-channels for each user can be assigned to one of the antennas. For the second
scenario, known as GAS, the number of RF chains is equal to the number of antennas in which each sub-channel for each user can be assigned to one of the antennas and subset of the antenna must be selected. It is shown that the resource allocation design can be formulated as a MOOP, which the conflicting objective functions are linearly combined into a single objective function employing the weighted sum method. The considered problem
MINLP is generally intractable. In order to obtain a computationally efficient suboptimal solution, the majorization minimization approach is proposed where a surrogate function serves as a lower bound of the objective function.~Simulation result demonstrated the superiority of the proposed method.~Furthermore, the proposed antenna selection scheme can strike an excellent balance of improving SE and EE.

\end{document}